# More Insight from Being More Focused

## Analysis of Clustered Market Apps

Maleknaz Nayebi
SEDS laboratory
University of Calgary
Calgary, Canada
mnayebi@ucalgary.ca

Homayoon Farrahi
SEDS laboratory
University of Calgary
Calgary, Canada
homayoon.farrahi@ucalgary.ca

Ada Lee
SEDS laboratory
University of Calgary
Calgary, Canada
ada.lee@ucalgary.ca

Henry Cho
Dept. of Engineering Science
University of Toronto
Toronto, Canada
henry.cho@mail.utoronto.ca

Guenther Ruhe
SEDS laboratory
University of Calgary
Calgary, Canada
ruhe@ucalgary.ca

## ABSTRACT

The increasing attraction of mobile apps has inspired researchers to analyze apps from different perspectives. As any software product, apps have different attributes such as size, content maturity, rating, category or number of downloads. Current research studies mostly consider sampling across all apps. This often results in comparisons of apps being quite different in nature and category (games compared with weather and calendar apps), also being different in size and complexity. Similar to proprietary software and web-based services, more specific results can be expected from looking at more homogeneous samples as they can be received as a result of applying clustering.

In this paper, we target homogeneous samples of apps to increase to degree of insight gained from analytics. As a proof-of-concept, we applied clustering technique DBSCAN and subsequent correlation analysis between app attributes for a set of 940 open source mobile apps from F-Droid. We showed that (i) clusters of apps with similar characteristics provided more insight compared to applying the same to the whole data and (ii) defining similarity of apps based on similarity of topics as created from topic modeling technique Latent Dirichlet Allocation does not significantly improve clustering results.

## 1. INTRODUCTION

Analytics of market apps is the emerging research direction with the ultimate goal of better understanding and influencing the development, evolution and market success of mobile apps [14]. Understanding and explaining why some apps are more important than others is one of the fundamental questions behind this whole effort. Data analytics is unlikely to give an ultimate answer on that, but likely to provide directions and triggers for improvement.

For collecting data, sampling is of key importance to achieve external validity of results [10]. While consideration of the quality of data is an increasing concern in (software engineering) data analytics, this paper wants to make the particular point of *not comparing apples with oranges*. Instead of making vague conclusions across a diverse set of apps, even if they have been selected randomly, one wants to detect patterns, relationships, rules or trends from a sample data set; the best chance for that is to have a set of comparable samples [13]. In other words, "one size does not fit all" not only applies to software methods, tools and techniques, but also applies to models and insights gained from data analytics in general and from the analytics of market apps in particular.

Clustering is an established methodology to create sets of items with similar characteristics. There are a variety of techniques aiming at grouping the original set of items into clusters of items being similar. The definition of similarity varies between the algorithms and the context that the analysis is applied to. For our proof-of-concept evaluation, we are using DBSCAN[3] as one out of the many possible techniques and tools.

For this paper we analyzed 940 apps. In particular, we were interested in analyzing factors correlated to market performance. We performed an analysis process including (i) finding relevant attributes to define clusters of interest, (ii) creating topics from the description of the apps and use these topics for clustering, (iii) performing the actual cluster analysis using DBSCAN, and (iv) applying a correlation analysis for both the original (raw) and the prepared (clustered) data sets and cmpare the results.

Main research questions (RQ's) to be answered are:

· **RQ1:** For the case of correlation analysis, does clustering raw data across all apps improve results?

- **RQ2:** For defining similarity, does inclusion of topics extracted from app descriptions improve clustering results (for correlation analysis)?

## 2. RELATED WORK

Data analysis and external validity of results heavily depend on the quality and size of the data used. Besides all the effort of data cleaning and preparation, we briefly discuss existing work of two approaches: (i) creating individual models for homogeneous subsets of the original set of all items, and (ii) proper sampling of data.

Analogy as a form of learning from similar cases has been widely used in software engineering. Analogy supports the transfer of patterns, best practices and lessons learned from one project or domain to other ones. Many effort estimation [9], cost estimation [5] and defect prediction models [12] were proposed based on the similarity and analogical reasoning. The critical point in using analogical reasoning is finding the relevancy between projects [6, 7]. Different clustering methods such as K-means, MESO, GAC [7], DBSCAN [8], FASTMap [11] and many others have been used to create more homogeneous data sets for reasoning. However, the use of market and software attributes were not discussed for mobile apps.

On the other side, Martin et al. [10] studied the effect of sampling bias on the results of app store mining with the focus on user reviews. In the context of mobile apps, k-means clustering was used in conjunction with topic modeling for detecting the abnormal behavior of apps [4]. More recently, Subaihin et al. [1] used hierarchical clustering to measure app similarity based on the textual description.

## 3. DATA GATHERING AND PRE PROCESSING

Our investigations were based on data from the F-Droid open source repository. The data collection process involved harvesting information from the market side and apps' development data. We started with gathering all the apps from F-Droid (1,844 apps as of May $20^{th}$, 2016). Looking into the source control repositories of these apps, we found that the majority of open source apps (73.9%) are hosted on GitHub. For all the apps from F-Droid, the HTML data from its respective F-Droid source was parsed to collect information regarding apps' *name, description, F-Droid link, source control repository, issue tracker* and *package name*. The apps were thereon filtered such that only apps with GitHub open source repositories remained. For the remaining apps, the *package names* obtained from F-Droid were used to access the app's webpage on GooglePlay. Here, we filtered out the apps that were not released into GooglePlay. For these apps - F-Droid apps on GitHub & GooglePlay, the HTML data from GooglePlay was parsed to collect information regarding the app.

We also gathered data from git repositories for each of the 940 apps in the sample using the GitHub URL available in F-Droid. The data extracted through such methods moved on for further analysis as to be detailed in later sections. The data gathering process is illustrated in Figure 1.

## 4. ATTRIBUTES USED FOR ANALYSIS OF MARKET APPS

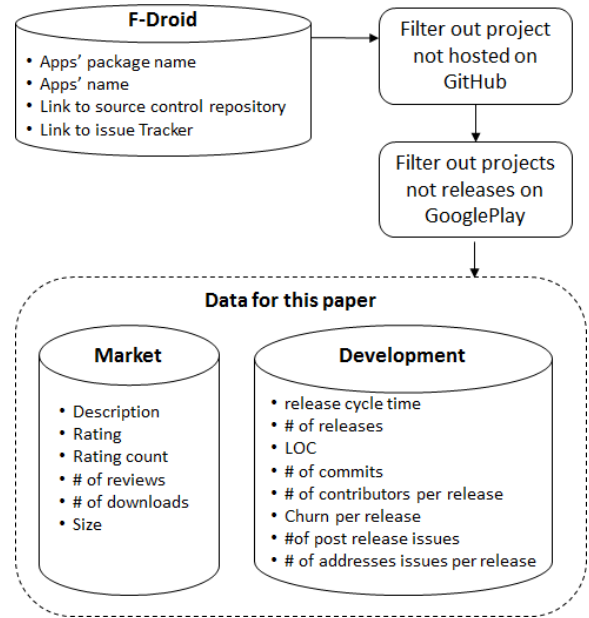

Figure 1: Data gathering process.

Among all the app data available in GooglePlay and GitHub, we focused on quantitative attributes complemented by the app description. For market attributes, these were the only ones available. For development attributes, we picked the ones that are potentially related to market performance. For example, we did not investigate on content maturity or pull requests. In what follows, we describe the different types of attributes in more detail.

### 4.1 Market Attributes

We gathered market information of each app from GooglePlay. Several information from GooglePlay including number of downloads, rating, number of reviews (same as number of raters), number of five star raters, number of one star raters, and size were collected. For the purpose of topic modeling, we also gathered the app descriptions.

#### 4.1.1 App Description

We used topic modeling on app descriptions. Topic modeling is the process of finding the topics of a large and possibly unorganized collection of documents. We used Latent Dirichlet Allocation (LDA) [2], for topic modeling. Topics describe the behaviour of apps and can be used to classify apps. The 940 F-Droid apps, were distributed across 41 different app store categories. However, only 15 categories had more than *one* app in them. So, following Gorla et al. [4] and also considering controllable clustering dimensions, we extracted 15 topics from app descriptions. The representative words for each topic are presented in Table 1. Each topic is represented by ten of the most represented stemmed words. For example, in Table 1 #4 presents *sharing* and #10 presents *network gaming*.

Once topic modeling was done, we performed a similarity analysis to see how similar an app description is to a topic. As a result of topic modeling, we assigned a probability vector *to each app*. Each item in this vector repre-sents the probability of assigning an app to a topic. The probability is in the range of [0,1]. Having 15 topics, each of the 940 apps have 15 items in the form of $app\#: <item1, item2, ..., item15>$ where # is the app id between 1 to 940. Also, $item1$ to $item15$ represents the similarity of $app$ to the mined topics $topic1, topic2, ..., topic15$. This vector is used as a clustering attribute.

**Table 1: Topics mined from F-Droid apps.**

| # | Representative stemmed words |
|---|---|
| 1 | avail, ad, support, permiss, install, user, access, connect, help, noti |
| 2 | deadlin, imperi, droidseri, wast, orgthi, loo, matric, wardrob, enterpri, blackberri |
| 3 | avail, simpl, text, device, app, file, inform, access, help, software |
| 4 | text, send, access, network, avail, card, view, button, email, github |
| 5 | read, ehlp, send, access, network, game, view, system, display, user |
| 6 | avail, text, email, help, simpl, inform, app, share, view, game |
| 7 | notif, help, download, game, search, share, require, bug, option, simpl |
| 8 | share, game, text, send, help, via, read, network, access, internet |
| 9 | avail, simpl, text, device, app, file, inform, access, help, software |
| 10 | network, game, access, download, read, send, share, select, display, ad |
| 11 | help, access, text, avail, search, user, color, github, email, bug |
| 12 | inform, avail, access, share, simpl, user, send, read, via, device |
| 13 | read, game, help, search, send, avail, ad, view, text, control |
| 14 | permiss, access, ad, file, user, avail, require, display, set, support |
| 15 | send, read, text, network, help, access, mode, user, ad, github |

**Table 2: Market and development attributes.**

| Context | Attributes (ID) |
|---|---|
| Market attributes | Rating (M1), # of raters (M2), # of five star raters (M3), #of one start raters (M4), # of downloads (M5), apk file size (M6). |
| Development attributes | Release count (D1), Lines of code (D2), Commit count (D3), Contributor count (D4), Release cycle time average (D5), Release cycle time variance (D6), Release churn average (D7), Release churn variance (D8), Release opened issues average (D9), Release opened issues variance (D10), release closed issues average (D11), release closed issues variance (D12). |

### 4.2 Development Attributes

We also gathered development data coming from git repositories. In particular, we mined release cycle time, number of releases, lines of code (LOC), number of commits, number of contributors per release, churn per release, number of post release issues, and number of addressed issues per release for all 940 apps.

The complete sets of *market attributes* and *development attributes* that were used for clustering are presented in Table 2.

## 5. METHODOLOGY

Our proposed methodology combines the functionality and strengths of various techniques. In what follows, we describe these components in a bit more detail.

### 5.1 Clustering

For a given set of items (apps in our case), clustering is a proven technique to determine subgroups of items having similar characteristics. There are a number of decisions to be made to actually perform clustering (e.g., the decision on the similarity measure or the attributes to be used for clustering). For our purposes, we mined combinations of several attributes being *rating*, *number of raters*, *number of five star raters*, *number of one start raters*, *number of downloads* and *size* of the app.

Having these attributes, **first** we selected all subsets of the set of all attributes to apply clustering. To find the best performing combination, having six market attributes, we performed $2^6 -1$ (= 63, the number of non-empty subsets) clusterings.

There are a variety of possible clustering techniques that could be taken. This study is not to compare them and to find out strengths and weaknesses of the individual techniques, but to show for one established technique (being DBSCAN [3]) the principal advantages for analyzing market apps. DBSCAN is a density-based clustering algorithm trying to group points with many nearby neighbors and marking points as outliers when they lie alone in low-density regions. DBSCAN has received substantial attention in theory and practice.

The *scikit-learn* library [15] was used for data pre-processing and clustering. We used this easy-to-learn and fast-to-use general purpose machine learning library in Python.

We used variance improvement to evaluate the quality of clustering. For a generated set of $n$ clusters, we used the following measures:

$$\text{Variance Improvement} = \sum_{i=1}^{n} \frac{(Var\_Cluster(i) - RawVariance)}{n * RawVariance}$$

**Var_Cluster(i)** describes the sum of all variance values between attributes defined resulting from looking at all apps being in cluster $i$.

**RawVariance** is the sum of all variances calculated from the raw data without clustering.

### 5.2 Brute-Force Correlation Analysis

Correlation analysis is a classical means to study the relationship between variables. Even though it does not allow causality conclusions, it is a first step to find more or less significant variables, e.g. for market analysis. To analyze the effect of clustering based on different combinations of

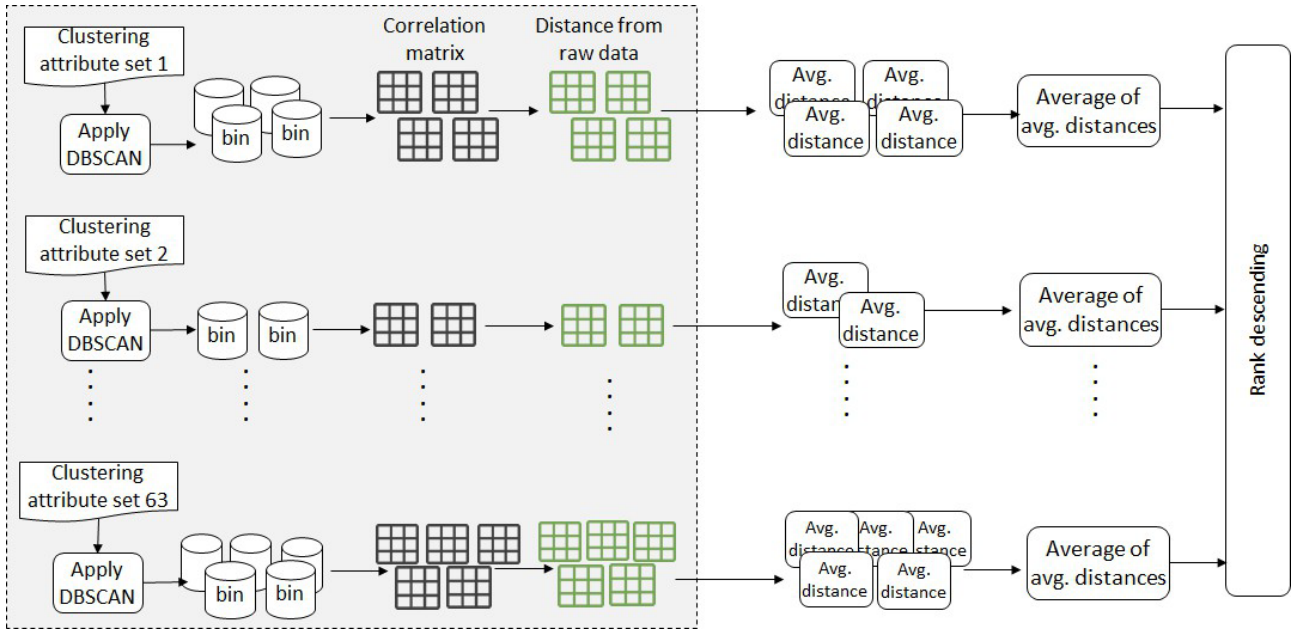

**Figure 2: Evaluation process based on average distance - The grey boxes show the core computation steps for our analysis.**

attributes, we performed an exhaustive *Brute-Force search*. This comprehensive search allowed us to find all the best clusterings when performing correlation analysis. For that purpose, the following process steps were taken:

**Step 1: Raw data correlation analysis** Calculate all correlations between all pairs of data attributes for the raw data (no clusters). While calculating these, we fill a symmetric matrix of correlation having the data attributes on columns and rows. For the six market attributes selected, this *baseline correlation matrix* is as below:

$$\begin{array}{c c c c c c c}
 & a_1 & a_2 & a_3 & a_4 & a_5 & a_6 \\
a_1 & 1 & * & * & * & * & * \\
a_2 & C_{a21} & 1 & * & * & * & * \\
a_3 & C_{a31} & C_{a32} & 1 & * & * & * \\
a_4 & C_{a41} & C_{a42} & C_{a43} & 1 & * & * \\
a_5 & C_{a5} & C_{a5} & C_{a5} & C_{a5} & 1 & * \\
a_6 & C_{a61} & C_{a62} & C_{a63} & C_{a64} & C_{a65} & 1 \\
\end{array}$$

In this (symmetric) matrix $c_{xy}$ shows the correlation between all pairs of attributes $x$ and $y$ in the raw data prior to clustering.

**Step 2: Attribute selection** Selection of all attributes to perform clustering.

**Step 3: Correlation analysis after clustering** Once the clustering for all subsets of attributes is preformed the symmetric correlation analysis we described in the Step 1, is calculated for all the clustreing bins.

**Step 4: Comparison** After calculating the correlation symmetric matrix for all clusters, we computed the *distances* between all raw data and clustering data correlation for all clusterings performed. The distance is defined as the difference between the coefficient after clustering minus the coefficient obtained from raw data (comparing Step 1 with Step 3). As the computational effort to perform Brute-Force correlation analysis is reasonable for our sets of attributes, its application guides the selection of the "best" clustering attributes for any correlation analysis to be done (with a fixed pair of attributes). As a result, the one having the maximum improvement against raw data coefficient is the recommended clustering.

For correlation analysis, in order to analyze the impact of clustering, we measured the distance of each cluster result from the baseline data. To do so, and following the process shown in Figure 2, we analyzed the correlation coefficients for all pairs of attributes across all the bins for a run of DBSCAN with each attribute set. First, we calculated the pairwise correlation between attributes in raw data and in each bin of each cluster. Next, we calculated the distance of pairwise correlations within a bin from the raw correlation. Then, we calculated the average correlation distance within each bin and then get the average across all bins of a cluster. Doing this for all clusterings with different subsets of attributes, we compared the performance of clusterings in terms of the difference they made from baseline.

## 5.3 Topic Modeling

LDA [2] was used to find topics in natural language text documents. Topics are created when words that tend to appear together frequently are found in the documents of the corpus. This method represents topics as probability distributions over the words in the corpus. We used LDA to extract topics (categorize) from app descriptions following the work of Gorla et al. [4]. We later used these topics for clustering the apps. To perform topic modeling, we first applied standard techniques of natural language processing

(NLP) for filtering and stemming. We filtered the app descriptions in four data pre-processing steps.

Firstly, unknown characters and non-English words were filtered out from the textual content. After, all the HTML tags were removed as they were not needed in the finding of topic models. Next, common English-language stop words such as "to", "a" and "is", were removed since they are not the focus of the analysis. Lastly, Porter's stemming algorithm [16] was used to map the words to their root form; for example, "playing" and "plays" would both be stemmed to their root word "plai".

After pre-processing of app descriptions, we used *Gensim* [17] to extract topics. Gensim is a robust library for unsupervised semantic modelling in Python and searches for similarities between strings of words, stylometry and term frequency as well.

Related to our former description of the steps of clustering based analysis, topic modeling is used here as a complementary way to create better clusters. As such, it enlarges the set of attributes elaborated in Step 2.

### 5.4 Tuning DBSCAN Parameters

As the first step and before the data can be fed into the DBSCAN algorithm, it has to be normalized by centering the data to the mean and scaling it to the unit variance. This ensures that the different attributes will have the same level of impact and are treated equally when analyzed in the algorithm.

DBSCAN uses two tuning parameters called *epsilon* referring to the neighborhood distance and *minPts* specifying the minimum number of neighbors. minPts is used to identify core, border and noise points. Points not assigned to a cluster were assigned to the noise bin. These parameters can be varied to divide all the data points into clusters (also called bins). Appropriate selection of these parameters depends on the nature of the data set and plays an important role in the quality of the clustering. To determine the proper parameter setting for our study, we ran several benchmarking tests combining and trying different epsilon and minPts values in conjunction. After considering the *market only attributes* we came up with the values $E$ = 0.05 and *minPts* = 15. A separate benchmarking was done considering *topic only attributes*. Selecting the parameters in the same manner as before, we ended up with $E$ = 0.9 and *minPts* = 15.

## 6. RESULTS

We used market attributes (M1 . . . M6) for performing the clustering and subsequently analyzed the relationship between pairs of market attributes and the pairs of development attributes. Separately, we also used 15 extracted topics for clustering and performed a correlation analysis on market and development attributes.

### 6.1 Effect of Clustering on Market and Development Attributes

We used market attributes (See Table 2) to perform clustering. We ran 63 clusterings by selecting all the non-empty subsets of six market attributes. To measure the impact of clustering, we performed pairwise correlation analysis between all market attributes and between all pairs of development attributes.

**Effect of clustering on market attributes:** Using market attributes for clustering, we followed the process illustrated in Figure 2. We first performed the core computation steps (gray area in Figure 2). Looking across all the clusterings, we presented the best achieved correlations (being closer to 0, 1, or -1) in Table 3. In this table, the items above the diagonal are the correlations in the raw data and the ones below are the best achieved correlations by clustering. From 63 runs of DBSCAN, the strongest impact has a distance of 0.181 and is the result of clustering based on *rating*, *number of five star raters*, and *number of one star raters*. This clustering performs almost equally to the clusterings based on:

- Rating, number of five star raters, number of one star raters, number of downloads, and
- Rating, number of five star raters, number of downloads.

The average distance created by each clustering is presented in Table 4. Running DBSCAN for 63 attribute combinations and across all bins (excluding noise bins), we got 1,741 defined pairwise comparisons (undefined correlation is the result of a zero standard deviation). Performing Brute-force correlation analysis showed that DBSCAN resulted in a substantial number of very high correlations being in the range of [0.95 1] (for 44.91% of the attribute pairs). The highly correlated samples can be used for training prediction models or to answer other specific research questions. For all the clustering iterations done, we measured variance improvement within each bin as discussed in Section 5.1. In 81.67% of the cases we observed variance improvements bigger than 99.0%.

**Effect of clustering on development attributes:** Running DBSCAN on 63 attribute sets all from market, we also investigate the similarities and diversities between development attributes for each app. In other words, we were interested to see to what extend market based clustering affects the heterogeneity of development attributes. We found that clustering based on *rating* and *binary size of an app* has the most impact on development attributes in terms of average distance of correlations. The second best impact is the clustering based on *rating, binary size of an app,* and *number of one star raters*. Table 4 demonstrates the top seven created distances and their respective attribute sets.

We also presented the best correlations achieved by clustering across all 63 runs in Table 6. In Table 6, the grey cells above the diagonal present the correlations in raw data while the below diagonal values are the best achieved correlations. Looking into the results of Brute-Force analysis across all bins created by all attribute subsets (8,645 pairwise defined correlations), DBSCAN could make low correlation (in the range of [0.05 0]) for 47.6% of pairwise correlations. For all the clusterings, we measured the improvement of variance within each bin as the average of the variance across all attributes; and in all of the cases, we observed variance improvements bigger than 99.0%.

**Table 3: Raw data correlations (above main diagonal)) and best appeared correlation after clustering between market attributes (below main diagonal).**

|       | $M_1$  | $M_2$  | $M_3$  | $M_4$  | $M_5$  | $M_6$  |
|-------|--------|--------|--------|--------|--------|--------|
| $M_1$ | 1      | -0.085 | -0.007 | -0.015 | -0.014 | 0.047  |
| $M_2$ | 0.000  | 1      | 0.998  | 0.978  | 0.977  | -0.008 |
| $M_3$ | -0.008 | 1      | 1      | 0.972  | 0.970  | -0.006 |
| $M_4$ | 0.003  | 0.999  | 0.999  | 1      | 0.997  | -0.009 |
| $M_5$ | 0.000  | 0.995  | 0.997  | 0.998  | 1      | -0.015 |
| $M_6$ | -0.029 | -0.004 | -0.007 | -0.001 | -0.000 | 1      |

**Table 4: Combination of market attributes creating highest average distance from baseline.**

| Rank | Attributes for Clustering | Avg. improvement |
|---|---|---|
| 1 | M1, M3, M4 | 0.181 |
| 2 | M1, M3, M4, M5 | 0.181 |
| 3 | M1, M3, M5 | 0.181 |
| 4 | M1, M3 | 0.178 |
| 5 | M1, M4, M5 | 0.168 |
| 6 | M1, M5 | 0.167 |
| 7 | M1, M4 | 0.162 |

**Table 5: Combination of development attributes creating highest average distance from baseline.**

| Rank | Attributes for Clustering | Avg. improvement |
|---|---|---|
| 1 | D1, D6 | 0.165 |
| 2 | D1, D4, D6 | 0.162 |
| 3 | D1, D4, D5, D6 | 0.159 |
| 4 | D1, D5, D6 | 0.154 |
| 5 | D5, D6 | 0.138 |
| 6 | D1, D3, D4, D5, D6 | 0.137 |
| 7 | D1, D3, D5, D6 | 0.134 |

**Table 6: Raw data correlations (above main diagonal) and best correlation after clustering between development attributes (below main diagonal).**

|  | $D_1$ | $D_2$ | $D_3$ | $D_4$ | $D_5$ | $D_6$ | $D_7$ | $D_8$ | $D_9$ | $D_{10}$ | $D_{11}$ | $D_{12}$ |
|---|---|---|---|---|---|---|---|---|---|---|---|---|
| $D_1$ | 1 | -0.008 | -0.007 | -0.015 | -0.014 | 0.047 | 0.026 | 0.007 | 0.033 | 0.051 | 0.032 | 0.040 |
| $D_2$ | 0.000 | 1 | 0.998 | 0.978 | 0.977 | -0.008 | 0.006 | -0.016 | 0.045 | 0.020 | 0.005 | 0.046 |
| $D_3$ | -0.0085 | 1 | 1 | 0.972 | 0.970 | -0.006 | 0.008 | -0.017 | 0.045 | 0.022 | 0.007 | 0.046 |
| $D_4$ | -0.002 | 0.999 | 0.999 | 1 | 0.997 | -0.009 | 0.004 | -0.012 | 0.050 | 0.018 | -0.000 | 0.050 |
| $D_5$ | 0.000 | 0.995 | 0.997 | 0.998 | 1 | -0.015 | 0.004 | -0.011 | 0.050 | 0.017 | -0.001 | 0.051 |
| $D_6$ | -0.029 | -0.004 | -0.007 | -0.001 | -0.000 | 1 | 0.078 | 0.008 | -0.016 | 0.073 | 0.094 | -0.013 |
| $D_7$ | -0.010 | -0.000 | 0.000 | 0.000 | 0.001 | -0.004 | 1 | 0.591 | 0.435 | 0.900 | 0.930 | 0.438 |
| $D_8$ | -0.000 | 0.001 | -0.003 | 0.003 | -0.003 | 0.006 | 0.810 | 1 | 0.348 | 0.268 | 0.316 | 0.286 |
| $D_9$ | 0.002 | 0.001 | 0.000 | 0.000 | 0.002 | -0.007 | 0.122 | 0.028 | 1 | 0.525 | 0.210 | 0.995 |
| $D_{10}$ | -0.013 | -0.002 | 0.003 | 0.000 | 0.006 | -0.006 | 0.969 | 0.008 | 0.723 | 1 | 0.916 | 0.567 |
| $D_{11}$ | 0.000 | 0.000 | 0.001 | 0.000 | -0.001 | 0.003 | 0.959 | -0.044 | 0.002 | 0.957 | 1 | 0.239 |
| $D_{12}$ | 0.002 | 0.001 | 0.000 | 0.001 | -0.001 | -0.005 | 0.131 | -0.046 | 0.998 | 0.755 | 0.001 | 1 |

## 6.2 Clustering in Consideration of Topics

In former studies [1, 4], topics were extracted topics from app descriptions to create latent clusters of apps. Inspired by that, we performed an exploratory analysis to compare the results of DBSCAN using market attributes with the results of DBSCAN using app topics related parameters. Applying LDA, we extracted 15 topics from the description of the apps and used the degree of apps' similarity to these topics as attributes for clustering with DBSCAN.

First, we performed clustering based on market attributes (M1 ... M6). Then, we analyzed the effect of clustering based on topic modeling on both market and development attributes. To perform the analysis, for each subset of attributes we performed the Brute-Force correlation analysis. Then, we got the minimum distance of each correlation from *zero*, *one* and *minus one* and considered the minimum value of these three. Then for each subset of attributes we calculated the summation for all bins. This measure indicates the effectiveness of a clustering in pushing attributes towards more isolation (zero correlation) or the strongest possible correlation (1 or -1). Then we got average across all the different runs of clusterings with different attribute subsets. The results of this comparison are presented in Table 7. As it is shown, using DBSACN (considering described set-tings) would perform much better in making homogeneous samples by using the market attributes instead of topics extracted from apps' description. To explore more, we used topic modeling in conjunction with market attributes for clustering and analyzed the impact based on average improvement (the same as previous section).

The highest average improvement for this clustering was 0.168 for market attributes and 0.166 for development attributes. Comparing these with the results of Table 4 and Table 5, the combination of market attributes with topics performed worse for market attributes. However, a very small (= 0.001) enhancement was observed for development attributes.

## 7. THREATS TO VALIDITY

The design and methodology of this study implies some threats to validity that needs a careful attention:

**Clustering method:** We selected DBSCAN for this study as one of the many clustering methods. However, our study did not include any evidence of the performance of this method in comparison to other clustering methods.

**Table 7: Comparison of clustering performance based on topics or market attributes.**

| Clustering Attributes | Analysis Domain | Avg. of all min. distances from {-1,0,1} |
|---|---|---|
| M1 ... M6 | M1 ... M6 | 1.239 |
| Topics | M1 ... M6 | 2.976 |
| M1 ... M6 | D1 ... D12 | 6.475 |
| Topics | D1 ... D12 | 24.851 |

**Tuning:** We performed a benchmark by systematically vary- ing $E$ and minPts and looked into the number and size of the clusters. We also implemented a k-distance diagram and got the peak of it to define the $E$. We ran a few studies consid- ering this optimal $E$. However, considering the low improve- ment of results in comparison to the non-optimal selection of $E$ and the computational complexity of dynamically search- ing for k-distance picks for all subsets of all attributes, we decided to pick one static MinPts and $E$ for all the subsets. One may expect to get slightly different results based on different tunings.

**Correlation analysis:** We used correlation as the simplest filter for the relation between two attributes. However, the analysis of correlation and the interpretation may impose internal validity to the study.

## 8. CONCLUSIONS

In the context of analyzing almost 1000 open source apps from F-Droid, and for the purpose of not comparing "ap- ples with oranges", we performed systematic clustering and studied its impact on the insights gained from correlation analysis. Our study showed that the highest distance from baseline data occurred in the clustering with the subset of three resp. four attributes. We could show that more insight can be expected from defining homogeneous subgroups of apps. However, the results need to be interpreted with care; no ultimate success or improvement can be easily recognized.

There is rich literature about analytical software engineering [18-45]. We see this research as the starting point for upcoming investigations and opportunities for looking into other analytics rather than just correlation. In addition, we need to find out if other clustering techniques might be more suit- able than DBSCAN which was used here for our purposes. Clearly, the approach appears to be more broadly applicable in data analytics of analyzing software repositories with samples coming from very different contexts.

## 9. ACKNOWLEDGEMENT

This research was partially supported by the Natural Sciences and Engineering Research Council of Canada, NSERC Discovery Grant 250343-12.

http://is.muni.cz/publication/884893/en.